\documentclass[english]{paper}
\usepackage[T1]{fontenc}
\usepackage[latin9]{inputenc}
\usepackage{geometry}
\usepackage{textcomp}
\usepackage{amsmath}
\usepackage{amssymb}
\usepackage{setspace}
\usepackage{babel}
\begin{document}
\begin{center}
\bigskip{}

\par\end{center}

\begin{center}
\textbf{Noiseless coding theorem proved by induction }
\par\end{center}

\begin{center}
\textbf{for finite stationary memoryless information sources }\vspace{1cm}

\par\end{center}

\begin{center}
Jozsef Szabo 
\par\end{center}

\begin{center}
Tampere University of Technology
\par\end{center}

\begin{center}
PL 553, 33101 Tampere, Finland
\par\end{center}

\begin{center}
jozsef.szabo@tut.fi
\par\end{center}

\begin{center}
5 July 2014
\par\end{center}

\begin{singlespace}
\begin{center}
\vspace{0cm}
\textbf{Abstract} 
\par\end{center}
\end{singlespace}

\begin{center}
\textit{Noiseless coding theorem for finite stationary memoryless
information sources is proved by using}
\par\end{center}

\begin{center}
\textit{induction on the number of source symbols and the inequality
of geometric and harmonic means.}
\par\end{center}

\noindent \textit{\bigskip{}
}\textbf{1.} \textbf{Introduction, terminology, notation}\vspace{0cm}

\noindent The noiseless coding theorem is well known since Shannon's
work ({[}S1{]}). The noiseless coding theorem for finite memoryless
stationary information sources is a special case of the general noiseless
coding theorem. It assumes that the information source is restricted
such that it outputs only symbols from a finite alphabet and each
outputted symbol has the same probability distribution(the source
is stationary) with no correlation between outputted symbols at different
time (the source is memoryless). The theorem gives a lower bound of
the average codeword length of any uniquely decipherable code used
for encoding the data coming from an information source with these
restricted properties.

\noindent Unlike the proofs using analysis ({[}MK6{]} and {[}R5{]})
here is given a proof using induction based on the number source symbols.\vspace{0cm}
We represent an information source (or shortly source) of this kind
$\,(S,P)$ (or shortly $\, S)$ where $\, S=\left\{ s_{1},...,s_{n}\right\} $
is the source alphabet and $\, P=\left\{ \left\{ p_{1},...p_{n}\right\} \right\} $,
$\sum_{i=1}^{n}p_{i}=1$, $\, p_{i}\epsilon\mathbb{R_{\text{+}}^{\star}}$,
where $\, p_{i}$ is the probability of outputting a source symbol
$\, s_{i},$ $\, i=1...n$.

\noindent \begin{flushleft}
We represent an $\, r$-ary code corresponding to source $S$ by $\,(C,f)$
(or shortly $\, C)$ where $\, C=\left\{ c_{1},...,c_{m}\right\} $,
$\, c_{j}\epsilon A^{*}$ ($\, A^{\star}=A^{0}\cup A^{1}\cup...$
is the Kleene closure), $\, j=1...m$, where $\, A=\left\{ 0,...,r-1\right\} $
an $\, r$-ary alphabet, r$\epsilon\mathbb{N}^{*}$ and $\, f:S->\wp(C)$$\setminus\left\{ \oslash\right\} $
is a function which for each source symbols defines the set of codewords
from which one can be used at any time to encode that source symbol.
It is usually assumed that $\, card\, f(s_{i})=1$, $\forall i=1...n$,
but that is not enforced by any means.
\par\end{flushleft}

\noindent \begin{flushleft}
For the code $\, C$ to be non-singular (non-ambiguous) we need to
assume$\, f(s_{i})\cap f(s_{j})=\oslash$, $\forall\, i\neq j$ where
$\, i,j=1...n$ . This implies that we have $\, card\, S\leq card\, C$
$(\, n\leq m)$.
\par\end{flushleft}

\noindent \begin{flushleft}
A non-singular code $\, C$ of $\, S$, where $\, card$$\, S=card\, C$$ $($\, card\, f(s_{i})=1$,
$\forall i=1...n$) is called uniquely decipherable if $\forall$w$\epsilon C^{\star}$
($\, C^{\star}$ is the Kleene closure) is a unique concatenation
of codewords from $\, C$ and is called instantaneous if non of the
codewords in the code are prefixes of each other. These concepts can
be extended to the case when $\, card$$\, S\leq card\, C$$ $ ($\, card\, f(s_{i})\geq1$,
$\forall i=1...n$). We could then define $\, C$ to be uniquely decipherable
if $\forall$w$\epsilon C^{\star}$ ($\, C^{\star}$ is the Kleene
closure) all the possible representation of $\, w$ as concatenation
of codewords correspond to(decodes to) a unique sequence of source
word $\, s$$\epsilon S^{\star}$; and we could define $\, C$ to
be instantaneous if non of the codewords in the code are prefixes
of each other(same definition as before).\vspace{0cm}

\par\end{flushleft}

\noindent \begin{flushleft}
Let us use the notation$\,\, H_{r}(S)$ for the $\, r$-ary entropy
(r$\epsilon\mathbb{N}^{*}$) of the finite stationary memoryless source
$(S,P)$. $H_{r}(S)=-\sum_{i=1}^{n}p_{i}\cdot log_{r}(p_{i})$, $\, n=card\, S$.
\par\end{flushleft}

\noindent \begin{flushleft}
Let us also use the notation $\,\, ACL_{r}(S,C)$ for the average
codeword length of any $\, r$-ary code $\, C$ of $\, S$. $ACL_{r}(S,C)=\sum_{i=1}^{n}p_{i}\cdot l_{i}$
when $\forall\, i=1...n:\, card\, f(s_{i})=1$ where $\, L=\left\{ \left\{ l_{1},...,l_{n}\right\} \right\} $
where $l_{i}=length(c_{i})$, $\, c_{i,}\epsilon f(s_{i}),\, i=1...n$.
\par\end{flushleft}

\noindent \begin{flushleft}
More generally if $\, card\, f(s_{i})\geq1$ we can say that $ACL_{r}(S,C)=\sum_{i=1}^{n}p_{i}\cdot(\sum_{u=1}^{card\, f(s_{i})}q_{i,u}\cdot l_{i,u})$,
when $\, q_{i,u}$ are well defined probabilities of encoding $\, s_{i}$
by $c_{i,u}\epsilon\, f(s_{i})$, $\sum_{u=1}^{card\, f(s_{i})}q_{i,u}=1,\, q_{i,u}>0,$
$\forall\, u=1...card\, f(s_{i})$ and $\, L=\left\{ \left\{ l_{1,1},...,l_{1,card\, f(s_{1})},...,l_{n,1},...,l_{n,card\, f(s_{n})}\right\} \right\} $
where $l_{i,u}=length(c_{i,u})$, $\, c_{i,u}\epsilon A^{\star},\, i=1...n;j=1...m$.\vspace{0cm}

\par\end{flushleft}

\noindent \begin{flushleft}
If any of the $\, q_{i,u}$ probabilities doesn't exist then average
codeword length cannot be defined as such. In that case, we can take
the sequence $ACL_{r,t}(S,C)=\cfrac{\sum_{i=1}^{n}(\sum_{z=1}^{t}\sum_{u=1}^{card\, f(s_{i})}k_{i,u,z}\cdot l_{i,u})}{t}$
of average codeword lengths after encoding t source symbols, $f_{i,t}=\sum_{z=1}^{t}\sum_{u=1}^{card\, f(s_{i})}k_{i,u,z}$
is the frequency of the source outputting $s_{i}$ when outputting
$\, t$ symbols and $\, k_{i,u,z}=1$ if the $\, z^{th}$ outputted
symbol is $\, s_{i}$ and is encoded to $\, c_{i,u}$ and otherwise
$\, k_{i,u,z}=0$. This sequence is kind of bounded below as follows:
\par\end{flushleft}

\begin{center}
$\, ACL_{r,t}(S,C)\geq\cfrac{\sum_{i=1}^{n}(\sum_{z=1}^{t}\sum_{u=1}^{card\, f(s_{i})}k_{i,u,z}\cdot min_{u=1...card\, f(s_{i})}l_{i,u})}{t}$
\par\end{center}

\begin{center}
$\,=$$\cfrac{\sum_{i=1}^{n}min_{u=1...card\, f(s_{i})}l_{i,u}\cdot(\sum_{z=1}^{t}\sum_{u=1}^{card\, f(s_{i})}k_{i,u,z})}{t}$
\par\end{center}

\begin{center}
$\,=$$\cfrac{\sum_{i=1}^{n}l_{i}\cdot(\sum_{z=1}^{t}\sum_{u=1}^{card\, f(s_{i})}k_{i,u,z})}{t}=$
\par\end{center}

\begin{center}
$\,=$$\cfrac{\sum_{i=1}^{n}f_{i,t}\cdot l_{i}}{t}=ACL_{r,t}(S,C^{'})$
, 
\par\end{center}

\begin{center}
where $(C^{'},g)$ is another encoding for the source $\, S$ where
$\, g(s_{i})=c_{i}\epsilon f(s_{i})$ for which $length(c_{i})=l_{i}=min_{u=1...card\, f(s_{i})}l_{i,u}$. 
\par\end{center}

\noindent It is clear that $lim_{t->\infty}ACL_{r,t}(S,C')=ACL_{r}(S,C^{'})$. 

\noindent So we have $ACL_{r,t}(S,C)\geq ACL_{r}(S,C^{'})$ if t is
big enough and so we could say formally$\, ACL_{r}(S,C)\geq ACL_{r}(S,C^{'})$,
even if it $ACL_{r,t}(S,C)$ doesn't converge to an exact value when
$\, t\rightarrow\infty$.\vspace{0cm}
Let us rely in our proof on the results obtained by Kraft ({[}K2{]}
and {[}R5{]}) and McMillan ({[}M4{]} and {[}R5{]}).

\noindent \begin{flushleft}
The proof that any extension of the source $\, S$ to the source$\, S^{p}$
etc. will not give better average codeword length either than the
entropy of $\, S$ can be revisited elsewhere ({[}R5{]}). This is
due to fact the property holds for $\, S$ (see our proof from 2.)
and from the fact that the source is assumed to be memoryless ({[}R5{]}).
\par\end{flushleft}

\noindent \vspace{0cm}
\textbf{2. Statement of the discrete noiseless coding theorem}

\bigskip{}

\noindent \textbf{Theorem:} If $\,(S,P)$ is a finite stationary memoryless
information source then $\, H_{r}(S)\leq ACL{}_{r}(S,C),$$\forall$$\, C$
uniquely decipherable code of $\, S$.

\vspace{0cm}

\noindent \begin{flushleft}
\textbf{Proof}.
\par\end{flushleft}

\vspace{0cm}

\noindent \begin{flushleft}
It is obvious that it is enough to consider only the case when $card\, C=card\, S$
($m=n$ ) without any loss of generality because for every code $(C,f)$
with $card\, C=m>n$ can always be constructed a code $(C^{'},g)$,
$card\, C^{\text{\textasciiacute}}=n$ for which holds $\,\, ACL_{r}(S,C^{'})\leq ACL_{r}(S,C)$.
This as for $\forall s_{i}\, i=1...n$ we can define for $\forall\, i=1...n$:
$\, g(s_{i})=\left\{ c_{i_{k}}\right\} ,$ where $\, f(s_{i})=\left\{ c_{i{}_{1}},...,\, c_{i{}_{v}}\right\} $
and $\, l_{i{}_{k}}=min_{u=1...v}\, l_{i_{u}},$ $\, l_{i_{u}}=length(c_{i_{u}}),$
$\, u=1...v$ (see also for details above in 1.). \vspace{0cm}

\par\end{flushleft}

\noindent \begin{flushleft}
If $\, r=1$ is obvious that $\, H_{r}(S)=-1\cdot log_{r}(1)=0$ since
the only possible uniquely decipherable codes are the form $\, C=\left\{ c_{1}\right\} $,
$c{}_{1}\epsilon A^{*}(A=\left\{ 0\right\} ).$ Equality would hold
only if $\, l_{1}=0$ which means $\, c_{1}=\lambda\epsilon A^{*}$
which is practically useless.
\par\end{flushleft}

\noindent \begin{flushleft}
Next on we assume $\, r\geq2.$\vspace{0cm}

\par\end{flushleft}

\noindent \begin{flushleft}
Due to Kraft's and McMillan's theorems it holds for every r-ary uniquely
decipherable code $\, C$ of source $\, S$ it is possible to construct
an instantaneous code $C^{'}$ of $\, S$ starting from the same codeword
lengths ($\, card\, C^{'}=card\, C$ and $\, length(c_{i}^{'})=length(c_{i}),\, i=1...n)$
$.$ This implies$ACL_{r}(S,C^{'})=ACL_{r}(S,C)$ and therefore without
loss of generality we can delimit ourselves to consider only instantaneous
codes. 
\par\end{flushleft}

\noindent \begin{flushleft}
Due to Kraft's theorem for every $\, r$-ary instantaneous code it
corresponds and $\, r$-ary tree and vice versa.\smallskip{}

\par\end{flushleft}

\noindent \begin{flushleft}
We will use induction over $\, card\, S$. To be able to apply the
induction step from a code $C$ of $S$ where $card\, C\leq n$ to
a code $C^{'}$of $S^{'}$ with $card\, C^{'}=n+1$ we will need to
do a reduction of a code of $\, n+1$ elements to a code of $\, n$
elements. This becomes possible if we could consider only such instantaneous
codes for which the corresponding $\, r$-ary tree doesn't have any
node(except the root) as standalone sibling node(standalone child
node). It is known that for every $\, r$-ary tree which does have
such nodes, by removing those we get an $\, r$-ary tree with smaller
average codeword length ({[}H3{]}, {[}R5{]}). \smallskip{}

\par\end{flushleft}

\noindent \begin{flushleft}
Finally, let us extend the set of special instantaneous codes for
which we want to do induction with 1 more element, the code $\, C_{\lambda}=\left\{ \lambda\right\} $
($\lambda\epsilon A^{0}$). This extension is required for the first
step of the induction. That doesn't restrict the generality anyhow
as we will prove the property for a bigger set than the special instantaneous
codes.
\par\end{flushleft}

\noindent \begin{flushleft}
\medskip{}

\par\end{flushleft}

\noindent If $\, card$$\, S=1$ then $\, P=\left\{ \left\{ p_{1}=1\right\} \right\} $
and $\, C=\left\{ \lambda\right\} $. So we have $\, H_{r}(S)=-p_{1}\cdot log_{r}(p_{1})=-1\cdot log_{r}(1)=0$,
$\, ACL{}_{r}(S)=p_{1}\cdot l_{1}=1\cdot0=0$. Here even equality
holds.\medskip{}

\noindent So we can do strong induction as any restricted instantaneous
code can be reduced to the code $C_{\lambda}=\left\{ \lambda\right\} $
ultimately. 

\noindent \begin{flushleft}
Let assume now that the inequality is true for any restricted and
extended instantaneous code $\, C$ (as above) of the source $(S,P)$
where $\, card\, C\leq n$, $\, n\geq1$ and let us prove it for a
code $\, C\text{\textasciiacute}$ of the source $\,(S^{'},P^{'})$
where $card\, C\text{\textasciiacute}=n+1$$ $. Let us use the notations:
$P\text{\textasciiacute}=\left\{ \left\{ p_{1},p_{2},\ldots,p_{n+1}\right\} \right\} $
and $\, C\text{\textasciiacute}=\left\{ x_{1},x_{2},\ldots,x_{n+1}\right\} $
relative to $\, S\text{\textasciiacute}$.
\par\end{flushleft}

\noindent \medskip{}
Due to the special properties of the instantaneous codes considered
there exists $\, x_{i_{1}},x_{i_{2}},...,x_{i_{s}}$, $2\leq s\leq r$,
where $\, length\, x_{i_{j}}=length\, x_{i_{k}}\forall\, j,k=1...s$
and the $\, x_{i_{k}}$-s, $\, k=1...s$ differ only in their last
symbol (to them corresponds sibling leafs in the corresponding $\, r$-ary
tree). \medskip{}
Let us reduce $(C^{'},P^{'})$ to $(C_{red},P_{red})$ where $C_{red}=C^{'}\cup\left\{ x_{red}\right\} $
$\setminus\left\{ x_{i_{1}},x_{i_{2}},...x_{i_{s}}\right\} $ and
the $\, x_{red}$ codeword is created by dropping the last symbol
from any of the codewords $x_{i_{k}}$ from $\, C^{'}$, $k=1...s$.

\noindent $\, P_{red}$ is formed by adding $p_{red}=p_{i_{1}}+...+p_{i_{s}}$
for $\, x_{red}$ and keeping the probabilities unchanged from $P^{'}$
for the retained codewords from C$^{'}$. 

\noindent We have:

\noindent \medskip{}

\begin{center}
$\, H_{r}(S\text{\textasciiacute})=H_{r}(S_{red})+p_{red}\cdot log_{r}(p_{red})-p_{i_{1}}\cdot log_{r}(p_{i_{1}})-p_{i_{2}}\cdot log_{r}(p_{i_{2}})-...-p_{i_{s}}\cdot log_{r}(p_{i_{s}})$
\par\end{center}

\begin{center}
$\, ACL{}_{r}(S^{'},C^{'})=ACL{}_{r}(S_{red},C_{red})-p_{red}\cdot l_{red}+p_{i_{1}}\cdot l_{i_{1}}+p_{i_{2}}\cdot l_{i_{2}}+...+p_{i_{s}}\cdot l_{i_{s}}$$\,=$ 
\par\end{center}

\begin{center}
$ACL\, r(S_{red},C_{red})-p_{red}\cdot l_{red}+p_{i_{1}}\cdot(l_{red}+1)+p_{i_{2}}\cdot(l_{red}+1)+...+p_{i_{s}}\cdot(l_{red}+1)$
$\,=$
\par\end{center}

\begin{center}
$ACL{}_{r}(S_{red},C_{red})+p_{red}\cdot l_{red}$ and by subtracting
we get: 
\par\end{center}

\begin{center}
$\, H_{r}(S\text{\textasciiacute)}-ACL{}_{r}(S^{'},C^{'})=$
\par\end{center}

\begin{center}
$H_{r}(S_{red})-ACL{}_{r}(S_{red},C_{red})+p_{red}\cdot log_{r}(p_{red})-p_{i_{1}}\cdot log_{r}(p_{i_{1}})-p_{i_{2}}\cdot log_{r}(p_{i_{2}})-...-p_{i_{s}}\cdot log_{r}(p_{i_{s}})-p_{red}$
\par\end{center}

\noindent As $card\, C_{red}\leq n$ then the induction hypothesis
is true for it: $\, H_{r}(S_{red})-ACL{}_{r}(S_{red},C_{red})\leq0$.\medskip{}

\noindent We only have to prove: 

\noindent \begin{center}
$p_{red}\cdot log_{r}(p_{red})-p_{i_{1}}\cdot log_{r}(p_{i_{1}})-p_{i_{2}}\cdot log_{r}(p_{i_{2}})-...-p_{i_{s}}\cdot log_{r}(p_{i_{s}})-p_{red}\leq0$ 
\par\end{center}

\noindent to end the proof. \medskip{}

\noindent By rearranging we get equivalently:

\begin{center}
$\, log_{r}(\cfrac{p_{red}}{r})^{p_{red}}$$\leq log(p_{i_{1}}^{p_{i_{1}}}\cdot p_{i_{2}}^{p_{i_{2}}}\cdot\cdot\cdot p_{i_{s}}^{p_{i_{s}}})$ 
\par\end{center}

\begin{center}
$\,\Leftrightarrow$ $\,(\cfrac{p_{red}}{r})^{p_{i_{1}}+p_{i_{2}}+...+p_{i_{s}}}\leq p_{i_{1}}^{p_{i_{1}}}\cdot p_{i_{2}}^{p_{i_{2}}}\cdot\cdot\cdot p_{i_{s}}^{p_{i_{s}}}$ 
\par\end{center}

\begin{center}
$\,\Leftrightarrow$ $\,(\cfrac{r\cdot p_{i_{1}}}{p_{red}})^{p_{i_{1}}}\cdot(\cfrac{r\cdot p_{i_{2}}}{p_{red}})^{p_{i_{2}}}\cdot\cdot\cdot(\cfrac{r\cdot p_{i_{s}}}{p_{red}})^{p_{i_{s}}}\geq1$$ $\medskip{}

\par\end{center}

\noindent Here we can assume first that $\, p_{i_{1},}p_{i_{2}},...,p_{i_{s}}\epsilon\mathbb{Q}$.
The general statement follows by continuity of the expression involved
and the density of $\,\mathbb{Q}$ in $\mathbb{R}$.

\noindent \medskip{}

\noindent Thus $\, p_{i_{k}}=\cfrac{f_{i_{k}}}{F}$, where $F=f_{i_{1}}+...+f_{i_{s}}$
and $f_{i_{k}}$ are positive integers, $k=1...s$. 

\noindent We need to prove that:

\begin{center}
$\,(\cfrac{r\cdot f_{i_{1}}}{f_{red}})^{f_{i_{1}}}\cdot(\cfrac{r\cdot f_{i_{2}}}{f_{red}})^{f_{i_{2}}}\cdot\cdot\cdot(\cfrac{r\cdot f{}_{i_{s}}}{f_{red}})^{fi_{s}}\geq1$$ $,
where $\, f_{red}=p_{red}\cdot F$.
\par\end{center}

\noindent Applying the inequality between geometric means and harmonic
means to the following sequence which has$\, F$ number of terms: 

\begin{center}
$\,$$\cfrac{r\cdot f_{i_{1}}}{f_{red}},..,\cfrac{r\cdot f_{i_{1}}}{f_{red}},\cfrac{r\cdot f_{i_{2}}}{f_{red}},...\cfrac{r\cdot fi_{2}}{f_{red}},...,\cfrac{r\cdot f_{i_{s}}}{f_{red}},...,\cfrac{r\cdot fi_{s}}{f_{red}}$$ $,
we get
\par\end{center}

\begin{center}
$\,\,(\cfrac{r\cdot f_{i_{1}}}{f_{red}})^{f_{i_{1}}}\cdot(\cfrac{r\cdot f_{i_{2}}}{f_{red}})^{f_{i_{2}}}\cdot\cdot\cdot(\cfrac{r\cdot f{}_{i_{s}}}{f_{red}})^{fi_{s}}=$
\par\end{center}

\begin{center}
$(\cfrac{r\cdot f_{i_{1}}}{f_{red}})\cdot\cdot\cdot(\cfrac{r\cdot f{}_{i_{1}}}{f_{red}})\,(\cfrac{r\cdot f_{i_{2}}}{f_{red}})\cdot\cdot\cdot(\cfrac{r\cdot f_{i_{2}}}{f_{red}})\cdot\cdot\cdot(\cfrac{r\cdot f{}_{i_{s}}}{f_{red}})\cdot\cdot\cdot(\cfrac{r\cdot f{}_{i_{s}}}{f_{red}})\geq$$ $
\par\end{center}

\begin{center}
$(\cfrac{f_{i_{1}}+f_{i_{2}}+...+f_{i_{s}}}{\cfrac{1}{\cfrac[l]{r\cdot f_{i_{1}}}{f_{red}}}+...+\cfrac{1}{\cfrac[l]{r\cdot f_{i_{1}}}{f_{red}}}+\cfrac{1}{\cfrac[l]{r\cdot f_{i_{2}}}{f_{red}}}+...+\cfrac{1}{\cfrac[l]{r\cdot f_{i_{2}}}{f_{red}}}+...+\cfrac{1}{\cfrac[l]{r\cdot f_{i_{s}}}{f_{red}}}+...+\cfrac{1}{\cfrac[l]{r\cdot f_{i_{s}}}{f_{red}}}})^{f_{i_{1}}+f_{i_{2}}+...+f_{i_{s}}}$ 
\par\end{center}

\begin{center}
$\,=$$\,(\cfrac{f_{red}}{\cfrac{f_{red}}{r\cdot f_{i_{1}}}\cdot f_{i_{1}}+\cfrac{f_{red}}{r\cdot f_{i_{2}}}\cdot f_{i_{2}}+...+\cfrac{f_{red}}{r\cdot f_{i_{s}}}\cdot f_{i_{s}}})^{f_{red}}$$ $
\par\end{center}

\begin{center}
$\,=(\cfrac{r}{s})^{f_{red}}\geq1$ , just what we wanted to prove.
\par\end{center}

\noindent Equality holds only if $\, p_{i_{1}}=...=p_{i_{s}}$ and
$\, s=r$.

\noindent It easy to see that via induction that equality holds only
if $\, p_{i_{k}}=\cfrac{1}{r^{l_{i_{k}}}}$, $\forall k=1,n$ where
$\, n=card\, S$ and also is of the form $\, n=z\cdot(r-1)+1$, $\forall$
$\, n\geq1$ where z is the number of internal nodes in the corresponding
$r$-ary tree.

\noindent If $\, card\, S=1$, form $\, H(S\text{)}=ACL{}_{r}(S,C)$
we don't get new information as it holds anyway. Since $P=\left\{ \left\{ p_{1}=1\right\} \right\} $
we have $p_{1}=\cfrac{1}{r^{0}}$ and $n=1=0\cdot(r-1)+1$ .

\noindent \begin{flushleft}
Let us use strong induction as above and retain the notations from
there.
\par\end{flushleft}

\noindent \begin{flushleft}
For $\,\, card\, S^{'}=n+1$ from $\, H(S\text{\textasciiacute)}=ACL{}_{r}(S^{'},C^{'})$
and doing same kind of reduction as for the proof above, we get:\medskip{}

\par\end{flushleft}

\noindent \begin{center}
$\, H_{r}(S\text{\textasciiacute)}-ACL{}_{r}(S^{'},C^{'})=$
\par\end{center}

\begin{center}
$H_{r}(S_{red})-ACL{}_{r}(S_{red},C_{red})+p_{red}\cdot log_{r}(p_{red})-p_{i_{1}}\cdot log_{r}(p_{i_{1}})-p_{i_{2}}\cdot log_{r}(p_{i_{2}})-...-p_{i_{s}}\cdot log_{r}(p_{i_{s}})-p_{red}\leq0$ 
\par\end{center}

\begin{center}
since we proved already that\medskip{}

\par\end{center}

\begin{center}
$\, H_{r}(S_{red})-ACL{}_{r}(S_{red},C_{red})\leq0$ and also that
\par\end{center}

\begin{center}
$p_{red}\cdot log_{r}(p_{red})-p_{i_{1}}\cdot log_{r}(p_{i_{1}})-p_{i_{2}}\cdot log_{r}(p_{i_{2}})-...-p_{i_{s}}\cdot log_{r}(p_{i_{s}})-p_{red}\leq0$
\par\end{center}

\begin{center}
it follows that both inequalities has to be equalities.
\par\end{center}

\noindent \begin{flushleft}
From $\, p_{red}\cdot log_{r}(p_{red})-p_{i_{1}}\cdot log_{r}(p_{i_{1}})-p_{i_{2}}\cdot log_{r}(p_{i_{2}})-...-p_{i_{s}}\cdot log_{r}(p_{i_{s}})-p_{red}=0$
it follows that $\, s=r$ and $\, p_{i_{j}}=p_{i_{k}},$$\forall j,k\leq s,$where
$\, p_{red}=p_{i_{1}}+...+p_{i_{s}}$ as we saw in the proof of the
theorem.
\par\end{flushleft}

\noindent \begin{flushleft}
From $H_{r}(S_{red})-ACL{}_{r}(S_{red},C_{red})=0$ and because $card\, S_{red}\leq n$
the induction assumption is true, so $\, p_{red}=\cfrac{1}{r^{l_{red}}}$
since $\, p_{v}=\cfrac{1}{r^{l_{v}}}$, $\forall v=1...card\, S_{red}$
and also holds by induction assumption that $\,\, card\, S_{red}=z\cdot(r-1)+1$
\medskip{}

\par\end{flushleft}

\noindent \begin{flushleft}
$\Rightarrow\, p_{red}=r\cdot p_{i_{j}}\Rightarrow p_{i_{j}}=\cfrac{p_{red}}{r}=\cfrac{1}{r^{l_{red}+1}}=\cfrac{1}{r^{l_{i_{j}}+1}}$,
$\forall j=1...r$ 
\par\end{flushleft}

\noindent \begin{flushleft}
as well as for the rest of the values $\, p_{v}=\cfrac{1}{r^{l_{v}}}$,
$\forall v=1...card\, S_{red},$$v\neq red$ which are inherited by
$\, S^{'}$ unchanged from $\, S$ $ $$\Rightarrow it$ is true for
all probabilities of $\, P^{'}$; additionally $\, card$$\, S^{'}=card\, S_{red}+r-1=z\cdot(r-1)+1+r-1=(z+1)\cdot(r-1)+1$
which end the proof. \medskip{}

\par\end{flushleft}

\noindent Let us observe that we got

\noindent \begin{center}
$\,(\cfrac{r\cdot p_{1}}{p_{1}+...+p_{s}})^{p_{1}}\cdot(\cfrac{r\cdot p_{2}}{p_{1}+...+p_{s}})^{p_{2}}\cdot\cdot\cdot(\cfrac{r\cdot p_{s}}{p_{1}+...+p_{s}})^{p_{s}}\geq1$,
$\,\forall s=1...r$ and $\forall p_{k}\mathbb{\epsilon\mathbb{R_{\text{+}}^{\star}}},\, k=1...s$ 
\par\end{center}

\noindent which is an interesting inequality as if we replace the
powers $\, p_{1}...p_{s}$ by $\,1$ and if $\, s=r$ then the reverse
inequality would hold,

\noindent \begin{center}
because by rearranging $\,(\cfrac{r\cdot p_{1}}{p_{1}+...+p_{r}})\cdot(\cfrac{r\cdot p_{2}}{p_{1}+...+p_{r}})\cdot\cdot\cdot(\cfrac{r\cdot p_{r}}{p_{1}+...+p_{r}})\leq1$
we get the known inequality between arithmetic and geometric means.\medskip{}

\par\end{center}

\noindent \begin{flushleft}
Additionally we found that: 
\par\end{flushleft}

\begin{center}
$\,(\cfrac{p_{1}+p_{2}+...+p_{s}}{r})^{p_{1}+p_{2}+...+p_{s}}\leq p_{1}^{p_{1}}\cdot p_{2}^{p_{2}}\cdot\cdot\cdot p_{s}^{p_{s}}$,
$\,\forall s=1...r$ and $\forall p_{k}\mathbb{\epsilon\mathbb{R_{\text{+}}^{\star}}},\, k=1...s$
and
\par\end{center}

\noindent \begin{center}
$p_{1}^{p_{1}}\cdot p_{2}^{p_{2}}\cdot\cdot\cdot p_{s}^{p_{s}}\geq\cfrac{1}{s},$
$\,\forall s\geq1$ integer and $\forall p_{k}\mathbb{\epsilon\mathbb{R_{\text{+}}^{\star}}},\, k=1...s$
where $p{}_{1}+...+p_{s}=1$. 
\par\end{center}

\noindent \medskip{}

\noindent \medskip{}

\noindent \begin{flushleft}
\textbf{References}\medskip{}

\par\end{flushleft}

{[}S1{]} Claude E. Shannon: A Mathematical Theory of Communication,
Bell System Technical Journal, 

Vol. 27, pp. 379\textendash{}423, 623\textendash{}656, 1948.

\medskip{}

{[}K2{]} L.G. Kraft, A Device for Quantizing, Grouping, and Coding
Amplitude Modulated Pulses, 

Q.S. Thesis, MIT, 1949.\medskip{}

{[}H3{]} Huffman, D. (1952). \textquotedbl{}A Method for the Construction
of Minimum-Redundancy Codes\textquotedbl{}. 

Proceedings of the IRE 40 (9): 1098\textendash{}1101. doi:10.1109/JRPROC.1952.273898

\medskip{}

{[}M4{]} B. McMillan, Two inequalities implied by unique decipherability, 

IRE Trans. Information Theory IT-2 (1956) 115-116\medskip{}

{[}R5{]} Steven Roman, Coding and Information Theory, Springer 1992.
\medskip{}

{[}MK6{]} David MacKay, Information Theory, Inference and Learning
Algorithms, 

Cambridge University Press 2003.
\end{document}